# Statistical Design of Thermal Protection System Using Physics-Informed Neural Network

Karthik Reddy Lyathakula[a], Aseem Muhammad[b], Sevki Cesmeci [c]

[a]Department of Mechanical and Aerospace Engineering, North Carolina State University, Raleigh, NC 27695, US
[b]Department of Mechanical and Materials Engineering, University of Cincinnati
[c]Assistant Professor; Department of Mechanical Engineering Georgia Southern University Statesboro, GA 30460 US

# Abstract

Thermal protection systems (TPS) of space vehicles are designed computationally rather than experimentally. They are validated using ground experiments, but all aspects of the flight cannot be replicated on ground. This ground-to-flight mapping introduces uncertainties which need to be accounted for while designing any thermal protection system. Thus, precise computational models along with uncertainty quantification in the models are required to design the TPS. The focus of this study is to estimate the thermal material parameters of TPS based on the target reliability requirements using statistical methods. To perform uncertainty quantification (UQ) of a system, a simulated model of the system needs to be solved many times on statistical samples, increasing the computational time and cost of the overall process. A physics-informed neural network (PINN) model is used in the analysis instead of traditional physics based numerical solutions. The accuracy of PINN is comparable to that of the numerical solution. To find the parameter distribution, sampling of the parameter space is performed using Sequential Monte-Carlo (SMC) method. The sampling method is efficient as it generates samples based on the target distribution in parallel and it also generates diverse samples for proper UQ. Combining the use of both PINN predictive model and SMC sampling, the framework can approximate the parameter distributions that satisfy the TPS design reliability constraints. The framework achieved remarkable increases in the speed of performing the reliability analysis of the TPS. This reliability analysis can be used for design optimization of the TPS based on risk analysis along with other systems of the vehicle.

# 1. Introduction

During the re-entry stage of a space vehicle, a significant amount of heat is generated due to air friction. This heat increases the temperature of the surface of the space vehicle, more in some areas than others. A heat shield or thermal protection system is designed to prevent any damage to the internal structure of the space vehicle. Different parts of the space vehicle will face different heat fluxes from the atmosphere and thus need to be designed accordingly. The goal of a successful TPS design is to prevent the internal components from reaching high "critical temperatures" while being as light as possible. This temperature is usually referred to as "back temperature" [1]. The design of TPS, however, faces multiple uncertainties caused by structural parameters, atmosphere variation, angle/speed of entry variation, material properties, modeling uncertainties etc. [2], [3]. This uncertainty needs to be accounted for in terms of the reliability of the system. Previous studies have highlighted the importance of considering uncertainties in TPS design [4], [5], [6], [7]. The failure of TPS can cause fatal accidents [8]. It is, thus, crucial for TPS to be successful despite these uncertainties. Reliability-based design optimization is the methodology in which a system is designed around target reliability score. It is widely considered as an advanced methodology that achieves optimal solutions while ensuring high system reliability [9], [10].

Extensive experimental testing is not possible on the ground since similar conditions are difficult to replicate. Thus, computational methods are used for designing the TPS. The reliability-based design optimization process includes two main steps: 1. Developing a predictive model that replicates the physics of the system. 2. Design optimization based on reliability constraints while accounting for uncertainties. The previous methods of design optimization mostly use physics-based tools such as finite element numerical solvers to simulate the re-entry phenomena [11]. The process of UQ and design optimization involves solving the model for many parameters sets [12]. This quickly becomes computationally expensive and time-consuming. A widely used approach is to create a surrogate model which can provide acceptable accuracy while increasing the computational efficiency [13], [14]. Surrogate modeling techniques can be performed using response surface methodology, statistical methods, and machine learning. Ravishankar employed the response surface method to conduct uncertainty analysis and predict the failure probability of thermal protection systems (TPSs) [15]. Tao utilized a statistical method, Kriging surrogate model, UQ of gas turbine blade end walls [16]. Guo optimized the parameters of metallic TPSs using a combination of Bayesian neural networks and genetic algorithms [17]. These surrogate models often require large amounts of data, and thus higher computational cost, especially when the training data is computationally expensive. The physics-informed neural network (PINN) has gained significant attention for its ability to approximate solutions to partial differential equations (PDEs) using neural networks in a direct and efficient manner [18]. Unlike traditional models, this technique combines data-driven methodology with the physics of the underlying system to build accurate and efficient surrogate models. This technique is widely used to develop fast and accurate surrogate models [19], [20], [21], [22].

Ever since the development of Apollo program and space shuttle, many researchers have studied the probabilistic design approach for TPS. The most common approach is the "worst-case analysis" which was used in the development of Apollo and space shuttle, due to which apollo was severely over-engineered

[23]. Another common method of determining the effects of data uncertainties is the use of "sensitivity analysis" [1]. Howel first utilized the Monte-Carlo method for uncertainties in thermal analysis [24]. Wright [25] and Ravishankar [15] also presented uncertainty analysis of aerothermal and thermal protection material responses using the Monte Carlo method. This technique provides a risk-based probabilistic analysis approach, whereby the TPS can be designed at the desired risk level, and the risk level can be combined with other components' risk level via system-level reliability analysis. Markov-chain Monte-Carlo (MCMC) method for efficient parameter generation is widely used in mathematical systems where target distribution is known [26]. MCMC is an 'intelligent' parameter sampling technique for efficient sampling. Sequential Monte-Carlo sampling uses MCMC as its foundation, but it is significantly faster than MCMC since it can generate parameters in parallel [27]. SMC has also been used for physical inverse modeling [28]. In the study, SMC along with a physics-based-neural-network predictive model is used.

## 2. Methodology

The methodology of the probabilistic design optimization process is as follows:

1. **Set Up a Numerical Model:** Build a mathematical model that represents how the thermal protection system (TPS) works and apply the necessary conditions at the edges i.e. critical temperature for the bottom side of the TPS.
2. **Create a PINN Surrogate Model:** Design a neural network that incorporates the rules and equations from the numerical model to guide its learning process.
3. **Train and Test the PINN Model**: Train the numerical model to make accurate predictions and compare its results with test data to ensure it works correctly.
4. **Define Limits and Reliability Goals:** Set the maximum allowable temperature critical temperature on the bottom of the TPS, decide on a reliability score to measure success, and define initial guesses for the key parameters to be optimized.
5. **Use MCMC for Parameter Sampling:** Apply Markov Chain Monte Carlo (MCMC) to explore different combinations of the key parameters. Keep sampling until the parameters stabilize and meet the critical temperature constraint.
6. **Analyze Uncertainty and Reliability:** Study the range of possible values for the key parameters and calculate how reliable the design is based on the results.
7. **Perform Sensitivity Analysis:** Test how changes in different parameters affect the model. Adjust the parameters and constraints to improve performance if needed.
8. **Finalize the Design:** If the parameters meet the design goals and stay within the constraints, stop the process. If not, continue adjusting and testing.

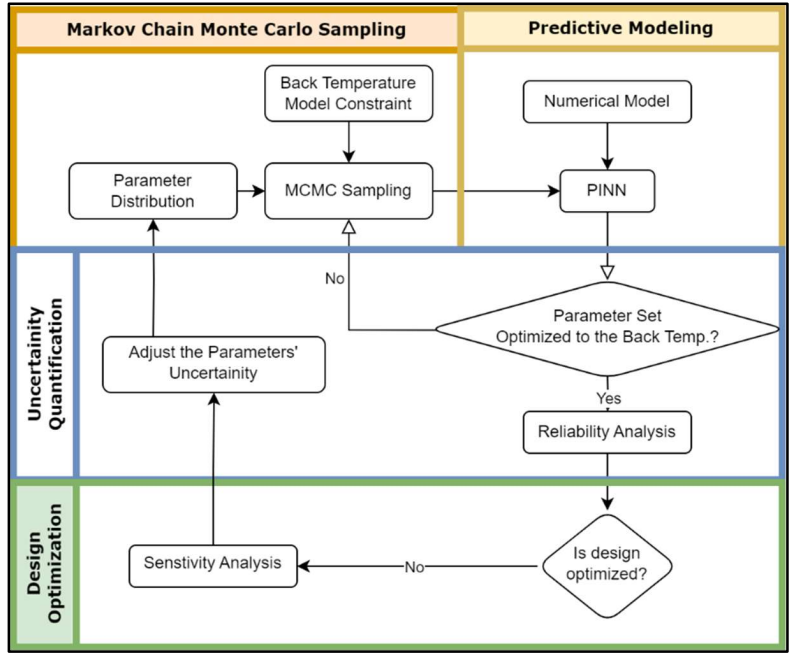

Figure 1: Probabilistic Design Process

## 3. Predictive Physics Model

The TPS of space vehicles experiences extreme heat flux during re-entry because of the air friction in the lower atmosphere. The back temperature should not exceed a certain limit on the inside of the TPS. For this study, heat transfer by radiation as well as heat transfer in longitudinal directions will be ignored. Thus, the simplified thermal response of a TPS can be modeled using the one-dimensional heat conduction equation. This partial differential equation describes how heat transfers through the TPS material over time:

$$\frac{\partial T}{\partial t} = \frac{k}{\rho c_p} \frac{\partial^2 T}{\partial x^2} \qquad 1$$

where $\rho$ is the material density, $c_p$ the specific heat capacity, T is the temperature, t is time, x is the spatial coordinate, and k is the thermal conductivity. The boundary conditions for this system include the applied heat flux at the outer surface and the insulated condition at the bottom surface:

On the inside i.e. at x=0, adiabatic boundary condition is applied:

$$\frac{\partial T}{\partial x}\big|_{x=0} = 0$$

On the outside i.e. at x=L, heat flux boundary condition is applied:

$$\frac{\partial T}{\partial x}\bigg|_{x=L} = -\frac{Q}{k}$$

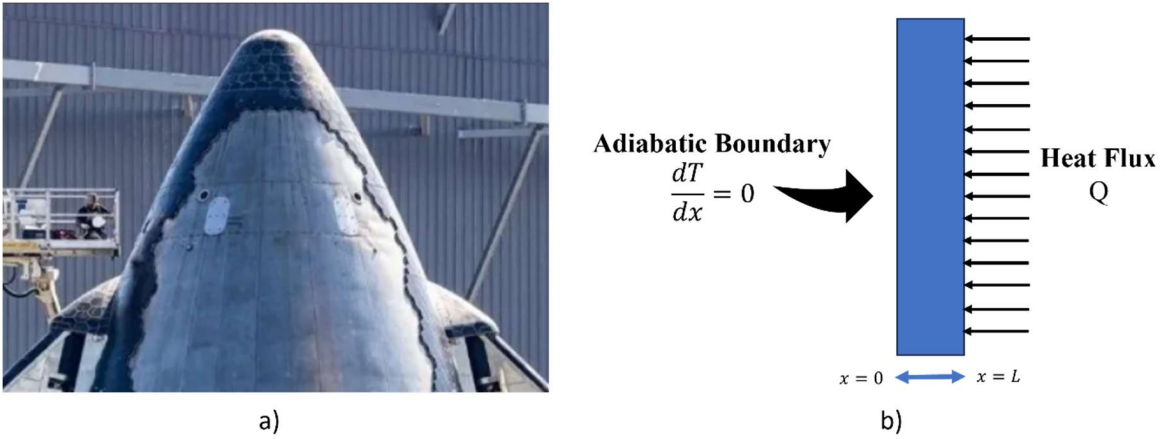

Figure 2: a) Thermal Protectional System of Space X's Starship [29] b) Thermal Model for TPS

For a TPS design to be successful, the back temperature $T_{back}$ at the bottom surface after the re-entry i.e. after time $t_{final}$ must remain below a critical temperature $T_{critical}$. In the TPS design, we define the constraint as such:

$$T_{back} < T_{critical}$$

$T_{back}$ is the final back temperature of the TPS after time $t_{final}$ while $T_{critical}$ is the temperature threshold for the final back temperature. Thus, the goal is to find the distribution of parameters that satisfy the design constraint.

### 3.1. Physics-Based Numerical Model

The physics-based numerical model solves the one-dimensional heat conduction equation using finite difference methods [30]. The spatial domain is discretized into N equally spaced nodes, and the temporal domain is divided into time steps determined by stability criteria. The governing equation is transformed into its discrete form:

$$T_i^{n+1} = T_i^n + \frac{k\,\Delta t}{\rho c_p \Delta x^2}(T_{i+1}^n - 2T_i^n + T_{i-1}^n)$$

Where $i$ represents the spatial node index and $n$ represents the time step. The above equation represents explicit numerical solver. An implicit solver was also implemented to compare the accuracy and stability of both solutions. Further details on implicit solution can be found from references [31].

## 3.2. Physics-Informed Neural Network

Deep learning has been successful in a variety of problems due to its ability of learn large amount of information and being scalable on large distributed hardware systems due to automatic differentiation [32], [33]. PINNs is a powerful deep learning framework for solving partial differential equations (PDEs) by embedding the governing physical laws directly into the neural network architecture. Recently, they have been widely used as a surrogate model for their accuracy and efficiency. PINNs incorporate physical constraints in the traditional machine learning models by incorporating the residuals of differential equations in the loss term of the neural networks [34]. Here, the PINN surrogate model for TPS is trained using the one-dimensional heat conduction equation stated above. The PINN architecture in the study uses thermal conductivity k, density $\rho$, specific heat capacity cp of the material, thickness of the TPS L, and time for the simulation. The neural network architecture is vanilla neural network with *softplus* as the activation function [35]. The loss function is the sum of initial condition loss, boundary condition loss, and physics loss.

Physics Loss:

$$f(x,t) = \frac{\partial u_\theta}{\partial t} - \frac{k}{\rho c_p} \frac{\partial^2 u_\theta}{\partial x^2} \qquad 2$$

$$L_{physi} = \frac{1}{N_{grid}} \sum_{i=1}^{N_{grid}} \left(f(x_i, t_i)\right)^2 \qquad 3$$

Where $N_{grid}$ is the number of points sampled in the space and time domain.

Initial Condition Loss:

$$L_{initial} = \frac{1}{N_{initial}} \sum_{i=1}^{N_{initial}} \left(u_\theta(x_i, 0) - u_o(x_i,)\right)^2 \qquad 4$$

Where $u_0$ is the initial condition provided. The initial condition in the case is set to 25 Celsius.

Boundary Condition Loss:

The boundary conditions are heating flux Q at x = L and insulation condition i.e. zero heat flux at x = 0. Thus, the boundary condition loss would be given by the sum of the two boundary condition losses as follows:

$$L_{boundary} = \frac{1}{N_b}\sum_{i=1}^{N_b}\left(\frac{u_\theta(0,t_i)}{\partial x} - 0\right)^2 + \sum_{i=1}^{N_b}\left(\frac{u_\theta(0,t_i)}{\partial x} + \frac{Q}{k}\right)^2 \qquad 5$$

Total Loss:

$$L_{total} = \alpha_1 L_{physi} + \alpha_2 L_{initial} + \alpha_3 L_{boundary} \qquad 6$$

Where $\alpha_1$, $\alpha_2$ and $\alpha_3$ are weights of their respective loss components. A visual representation of the PINN is given in the figure.

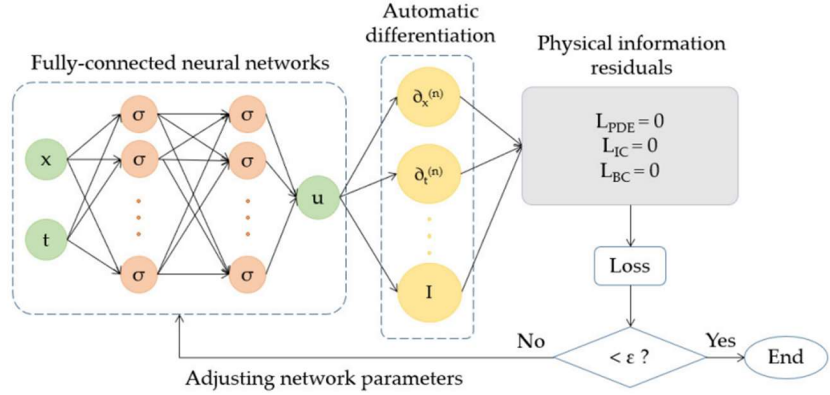

Figure 3: Schematic of a physics-informed neural network (PINN) [5]

# 4. Parameter Estimation and Uncertainty Quantification

Uncertainty quantification (UQ) in engineering systems typically relies on the Monte Carlo method, which follows a structured three-step approach [1]. The process begins with the selection of initial parameters, where geometric dimensions and material properties are chosen based on available materials or engineering estimates. This is followed by the crucial step of uncertainty characterization, where sources of uncertainty in parameters are identified, and appropriate probability distributions are assigned. Typically, these parameters are assumed to follow normal or uniform distributions with specified means and standard deviations. The third step involves predictive modeling, where simulations are performed on the sampled parameter sets. Once the results are obtained from the simulation, thermal reliability analysis is conducted alongside sensitivity and correlation analyses. The core methodology involves performing thermal analysis using multiple parameter sets drawn from assumed statistical distributions. Through statistical 'averaging' of these models, the total system uncertainty can be determined.

For estimation/optimization of the parameters, the process is repeated by changing the initial estimates in each iteration manually until the optimized parameters are achieved. The effectiveness of Monte Carlo analysis fundamentally depends on several critical factors: the accurate identification of uncertainty sources, appropriate selection of parameter ranges, and proper distribution assumptions. The parameter estimation is then based on successful execution of all the above coupled with iterative refinement of the parameters. The system's design process concludes once it meets the reliability requirements, such as the back temperature in thermal protection systems (TPS).

However, this method faces significant challenges. One major difficulty lies in accurately estimating initial parameter distribution. Incorrect distribution choices can result in suboptimal designs or, in worst-case scenarios, compromised system performance. Second, the process requires manual optimizing the parameter sets for the optimal design. To mitigate these risks, normal distributions with large standard deviations are utilized, necessitating numerous simulations. This approach becomes particularly problematic when each simulation requires computationally expensive predictive models, substantially increasing the overall computational burden.

## 4.1. Bayesian Inference

The Bayes' theorem of inverse probability is used to update the uncertain parameter set (posterior distribution), material properties and sizing parameters, based on the prior distribution of the parameter set and the likelihood:

$$\pi(\theta|D) = \frac{\pi(D|\theta)\pi(\theta)}{\pi(D)} \qquad 7$$

where θ represents the parameter set $[k, rho, c_p]$ and (D) represents the design constraints, $P(D|\theta)$ is the likelihood function. The prior distribution $\pi(\theta)$ is the initial distribution of the parameter set $\theta$. $\pi(D)$ is the normalization factor. The Bayes' theorem expresses the posterior distributions $\pi(\theta|D)$ as a function of likelihood $\pi(D|\theta)$ and prior distribution $\pi(\theta)$.

It is an assumption that the error between $T_{back}$ predicted from the prior distribution $\pi(\theta)$ of parameters and $T_{critical}$ follows a gaussian distribution with standard deviation $\sigma$. As such, the likelihood function is obtained as follows:

$$\pi(D|\theta) = \frac{1}{(2\pi\sigma^2)^{n/2}} \exp(\frac{(T_{critical} - T_{back})^2}{2\sigma^2})$$

Typically, the prior would be calculated from the assumed distribution around the known/estimated values of $\theta$. In the case, it could be the values of materials and initial sizing. It can be written as:

$$\pi(\theta) = \pi(\rho) . \pi(c_p) . \pi(k) \text{ where } \theta = [k, rho, c_p]$$

Where these distributions could normal or uniform or some other distribution around an estimated mean and available standard deviations. However, we do not use these assumed distributions for repeated sampling and rather use MCMC for sampling, which will be discussed in the next section. We only use these for sampling initial parameter set $\theta$ from an initial normal distribution.

The normalization constant $\pi(D)$ is calculated as the integral of the likelihood and prior over the complete range.

$$\pi(D) = \int \pi(D|\theta)\pi(\theta)\, d\theta$$

The posterior distribution $\pi(\theta|D)$ reflects the optimal parameter values after considering both prior and the critical temperature model constraint.

In theory, Bayesian inference can help us in parameter estimation by using prior estimation of the parameters and improving based on the physical model. However, practically, there are a few challenges. One, the calculation of the normalization constant $\pi(D)$ requires integrating over all possible parameter values, which is often computationally infeasible, especially in high-dimensional spaces. Two, the posterior distribution is highly nonlinear and is difficult to estimate using standard numerical techniques. To quantify the uncertainty, the normalization constant in the posterior needs to be solved. The parameter posterior is a highly nonlinear equation and is difficult to estimate using standard numerical techniques. For large-scale models or high-dimensional parameter spaces, Bayesian UQ can become prohibitively expensive due to the need for thousands of model evaluations. Another issue is that even though it updates the posterior based on the design constraints, the choice of prior still influences the posterior significantly. In practice, Bayesian methods can be used with efficient sampling methods such as Markov Chain Monte-Carlo.

## 4.2. Markov Chain Monte-Carlo

The Markov Chain Monte Carlo (MCMC) method provides an efficient approach for sampling from complex probability distributions, particularly useful in the context of TPS design. While traditional Monte Carlo methods sample independently from assumed distributions, MCMC generates a sequence of dependent samples that converge to the desired target distribution. This is particularly valuable when the target distribution is complex, as is often the case in Bayesian inference.

Metropolis-hasting algorithm is often utilized to implement MCMC [36]. The target distribution $\pi(\theta|D)$ is defined based on the critical temperature constraint $T_{critical}$ and a reliability score $R$ as given in the eq. x. We assume that the target distribution $\pi(\theta|D)$ is a gaussian distribution which achieves a reliability of R with a standard deviation of $\sigma_{target}$. With this assumption, reliability score $R$ is used to find the z-score of the gaussian distribution, which in turn, is used to find the mean of the distribution. The distribution can then be generated from the mean $\mu_{target}$ and $\sigma_{target}$. The reliability score is defined as:

$$R = P(T_{back} \leq T_{critical}) = N\left(\frac{T_{critical} - \mu_{target}}{\sigma_{target}}\right) \quad \quad 8$$

Once a target distribution is available, the chain is initialized with a parameter set $\theta^{m=0}$ drawn from the prior distributions $\pi(\theta)$. At each iteration, a new candidate sample $\theta^*$ is proposed using a proposal distribution $q(\theta^*|\theta^m)$. The proposal distribution $q(\theta^*|\theta)$ is a multivariate symmetric gaussian distribution centered around the previous sample $\theta^m$:

$$q(\theta^*|\theta^m) = N(\theta^m, \Sigma)$$

Where $\Sigma^m$ is the covariance matrix of the parameter set. The temperature $T_{back}$ is calculated using the parameter sample $\theta^m$. A ratio between the likelihood of the previous sample $\theta$ and the new sample $\theta^*$ is calculated as:

$$r = \frac{\pi(\theta^*|D)}{\pi(\theta|D)} = \frac{\pi(D|\theta^*)\pi(\theta^*)}{\pi(D|\theta)\pi(\theta)}$$

The benefit of using MCMC is that instead of the normalization function P(D), doesn't need to be explicitly calculated because of the ratio. Now, the candidate sample $\theta^*$ is accepted or rejected based on a random number called acceptance ratio $\alpha$. It can be sampled from a uniform distribution from 0 to 1 for MCMC reasonable sampling. If the ratio $r$ calculated is more than the acceptance ratio , the candidate sample is used, and then next iteration begins with this sample being $\theta_t$. This process is repeated until a stationary state is reached or a specified number of times. In the case, 1000 samples are generated from 3 separate chains (starting from different initial points). All the samples, even the burn-in samples, are included in the uncertainty quantification process. It is also made sure that the chain reaches stationary state and explores the entire parameter space.

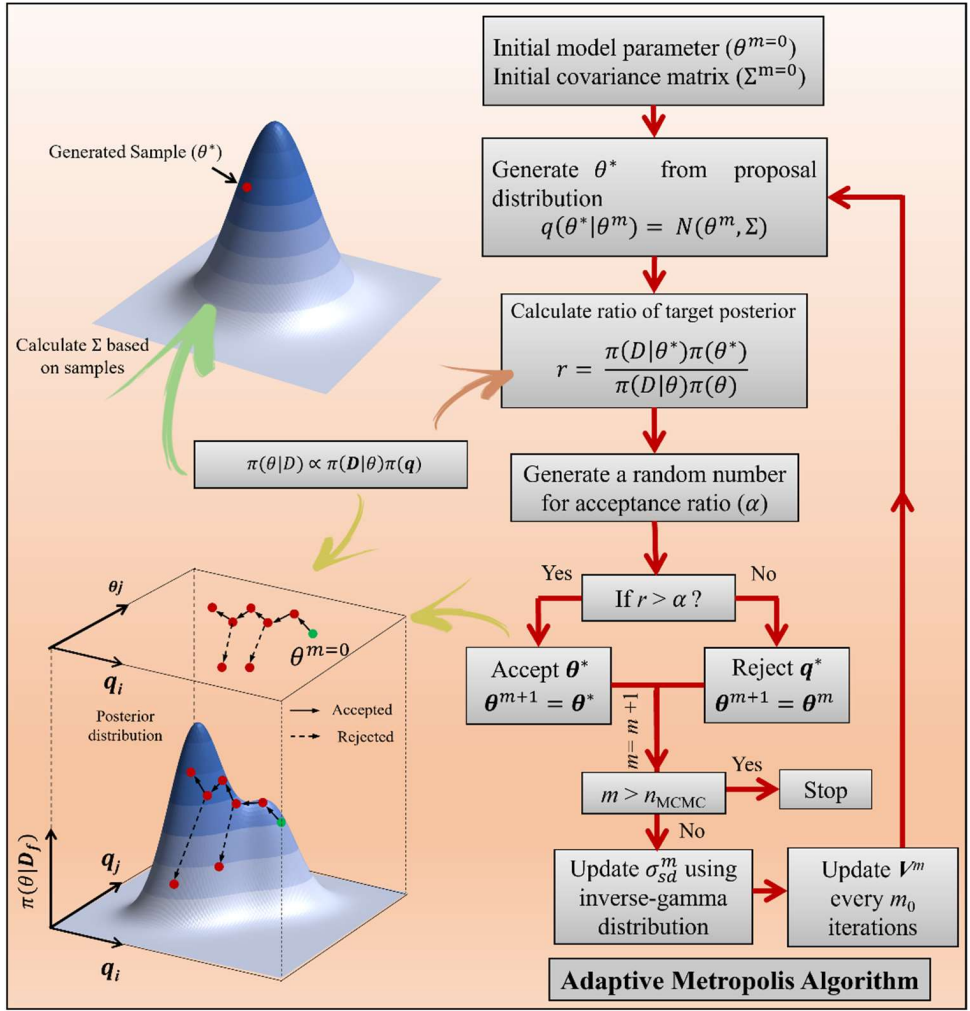

*Figure 4: Markov Chain Monte-Carlo Process Diagram[37]*

## 4.3. Sequential Monte-Carlo

While MCMC provides robust sampling, its inherently serial nature limits computational efficiency for complex TPS designs. Secondly, it can be entrapped within the local modes and never reach the stationary state [27]. The Sequential Monte Carlo (SMC) method overcomes these limitation through parallelized sampling while maintaining accuracy in posterior approximation [38].

The SMC method generates samples through a sequence of tempered distributions, where the target distribution $\pi(\theta|D)$ is approached gradually through intermediate distributions $\pi_t(\theta|D)$:

$$\pi_t(\theta|D) \propto \pi(D|\theta)\,\phi_t \pi(\theta)$$

where $\phi_t$ is the tempering parameter, which transitions from $\phi_0 = 0$ (corresponding to the prior distribution $\pi(\theta)$ to $\phi_{nSMC} = 1$ (the posterior distribution $\pi(\theta|D)$)

The method initializes by generating $N$ particles $\theta$ from the prior distribution $\pi(\theta)$. Each particle is assigned an initial weight $w_i^0 = 1/N$. The weights are normalized such that the sum of all weights is equal to one. At each intermediate step $t$, the weights of the particles are updated based on the tempered likelihood ratio:

$$w_i^t = w_i^{t-1} \cdot \frac{\pi(D|\theta_i)^{\Delta\phi_t}}{\sum_{j=1}^{N} w_j^{t-1} \cdot \pi(D|\theta_j)^{\Delta\phi_t}}$$

Where $\Delta\phi_t = \phi_t - \phi_{t-1}$. This ensures that the particles reflect the new tempered distribution $\pi_t(\theta|D)$.

To avoid particle degeneracy, where most particle weights become negligible, resampling is performed when the Effective Sample Size (ESS) drops below a predefined threshold $ESS_{thre}$. The $ESS_t$ is computed as:

$$ESS_t = \frac{(\sum_{n=1}^{N} W_t^n)^2}{\sum_{n=1}^{N} (W_t^n)^2}$$

The resampling for each sample is performed using MCMC kernel in parallel. $\pi_t(\theta|D)$ is used for initial sampling. Acceptance ratio $\alpha$ is used for acceptance/rejection of the samples until we get N samples that satisfy the acceptance ratio. Detailed process is described in the figure.

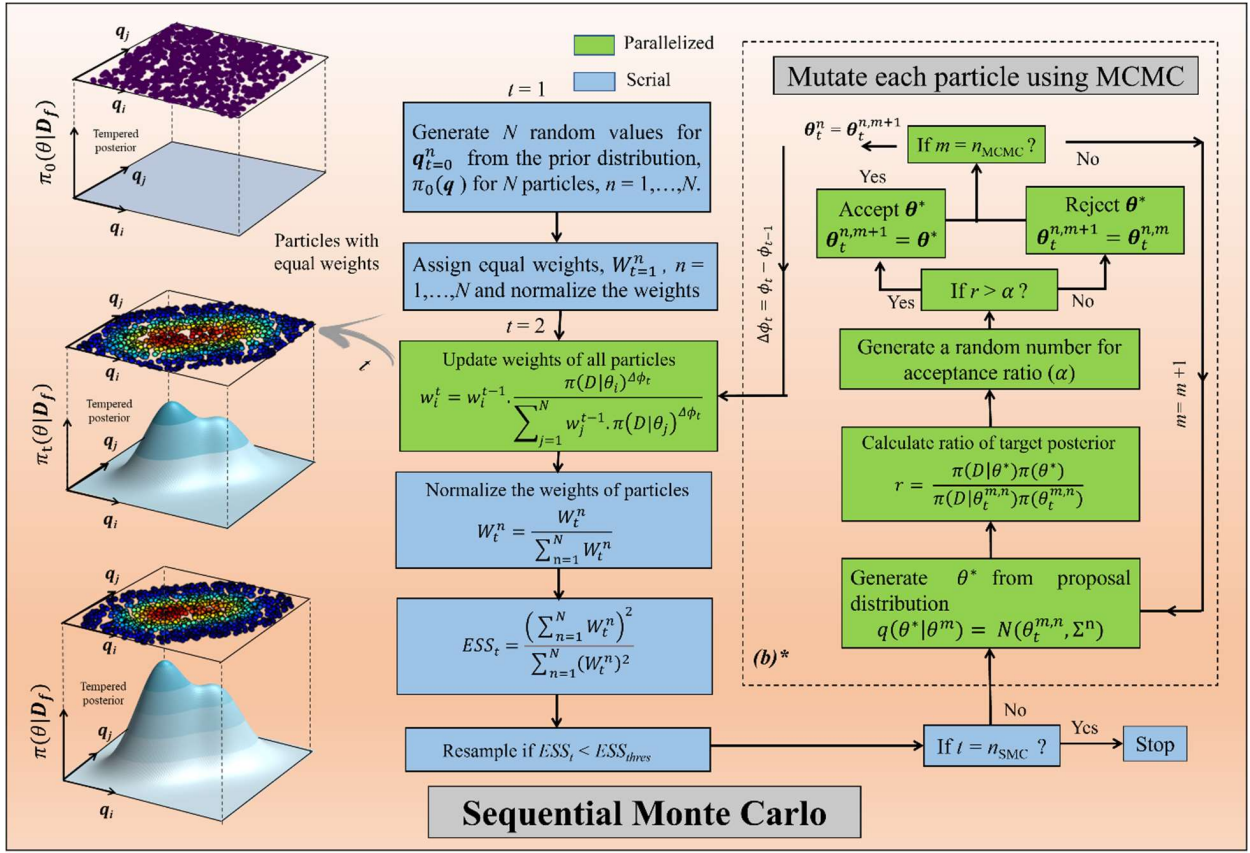

*Figure 5: Sequential Monte Carlo Process [37]*

# 5. Results and Discussion

This section presents the results of the methodology presented in section 2. The numerical solution of the Eq. 2 is implemented in Python programming language using finite difference method (FDM) technique. In this study, the heat flux is set to 10,000 $W.m^{-2}$ and the thickness of TPS is set to 7mm. These parameter values are chosen based on typical values of the thickness of TPS and the heat flux during the re-entry stage. The heat flux and the thickness remain constant for the study while the material properties density $\rho$, thermal conductivity $k$ and heat capacity $c_p$ will be optimized. A RCC composite AS4/3501-6 was chosen as a potential material for TPS material [39]. This material's is chosen as the validation case for the validation of numerical and PINN solution. The parameters for the material are given in table x.

| Parameter | Symbol | Value |
|---|---|---|
| Specific Heat Capacity | $c_p$ | 1050 J/kg K |
| Thermal Conductivity | $k$ | 0.65 W/m K |
| Density | $\rho$ | 1509 kg/m3 |

The numerical solution is implemented using both implicit and explicit differential solvers. A grid convergence study was performed to compare the accuracy of the results at different meshing levels. It is found that 100 grid points in the space domain i.e. x domain is sufficient. For time domain, the level of discretization depends on the solver. For implicit solver, the number of time steps is the same as the number of points in the space domain. For explicit solutions, CFL number was set to 0.2 for numerical stability.

The temperature, length and time domains are normalized to improve the numerical stability of both numerical solutions and PINN. The normalization is performed according to the Eq x. The choice of $T_{norm}$ is arbitrary and should be based on the scale of the temperature. In this study, a value of 100 is used.

$$T' = \frac{T}{T_{norm}}, t' = \frac{t}{t_{final}}, x' = \frac{x}{L}$$

The inputs to PINN are the points in the physics domain, points on the boundary and initial conditions, and the parameters $[k, \rho, c_p]$. In this implementation, $\rho$ and $c_p$ are combined to find the thermal density ($\rho * c_p$) of the TPS which is then used as the input. The output is the temperature on the input points. The PINN can not only predict the back temperature on the bottom end i.e. $x = 0$, but it can also predict at the complete length domain i.e. $x = 0 \text{ to } L$. Figure x shows the training loss plot against the iteration number for a typical case. Even though the PINN performs well in a wide range of parameters, there is still a need to define a range which is used to train the PINN. The PINN performs best in the parameter range which was used in the training process. We use a wide range centered around the properties of RCC material. Random sampling from a uniform distribution is performed within that range for training of PINN. The PINN surrogate model presented was implemented using PINA (Physics-Informed Neural network for Advanced Modeling) library built on top of Pytorch. The neural network (NN) architecture is a fully connected neural network with 3 hidden layers, each containing 30 neurons. To ensure smooth high-order derivatives while maintaining non-negative outputs, the softplus activation function was selected. The training process utilizes randomly selected 100 points for the physics domain, and 100 points combined for the boundary and the initial condition. Training was conducted using the Adam optimizer with 0.006 learning rate for 2000 epochs. The process was executed on both a CPU (13th Gen Intel(R) Core (TM) i5-13420H) and a

GPU (NVIDIA GeForce RTX 4050 Laptop GPU) to compare the performance. The SMC implementation was performed using the package SMCPy from NASA [40].

PINN produced a root-mean-square error of 3.43 °C for the complete time and space domains. The maximum error was observed at the initial condition prediction, which is not significant to the analysis. It is observed during the training process that the error does not compound for longer simulations. The numerical solution, the PINN predicted solution and the absolute error between them is provided in the figure below. The findings demonstrate that PINN can produce accurate results reliably.

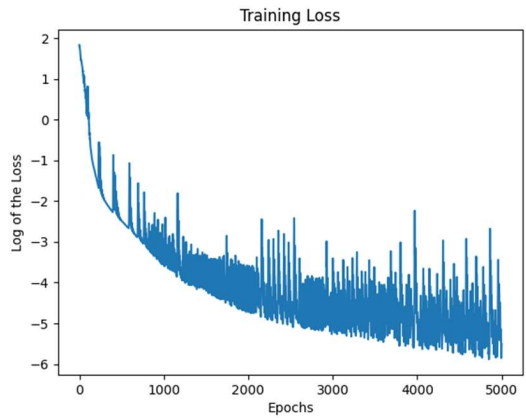

Figure 6: Loss Curve of PINN for the validation case

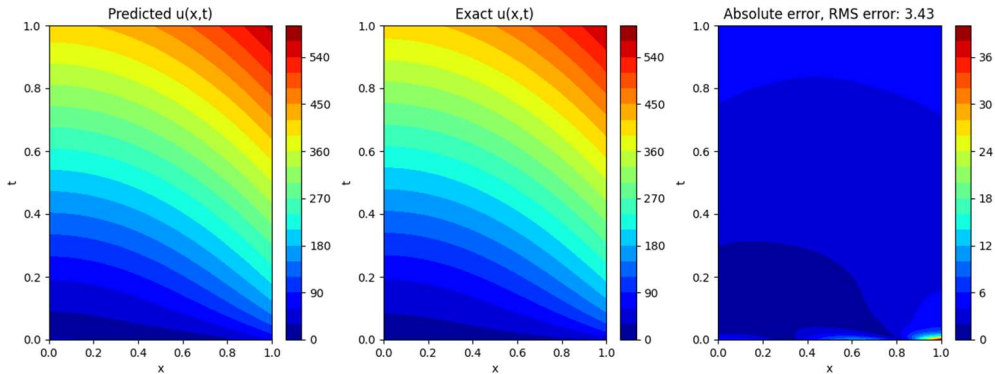

Figure 7: a) PINN Prediction b) Numerical Solution c) Absolute Error between PINN and Numerical Solution

Once the PINN is trained, it can be utilized to predict the temperature field for any parameter set θ. In this study, three target distributions of $T_{critical}$ are generated with different reliability scores. In practice, the reliability goals are defined according to overall system-level reliability goals. Here we use 95%, 99% and 99.999% and compare the parameter distributions. An initial random sample of $\theta$ is picked and SMC is used to generate further samples according to the target distributions. 10,000 samples are generated for each reliability score. The figures attached show a comparison of different target distributions and the parameters' distribution. It can be observed that the target distributions have lower mean back temperature as the reliability score increases, as expected. Figure 8c shows that thermal diffusivity needs to be higher for higher thermal conductivities for the same reliability score. One noticeable observation in Figure 8d is that there is a sharp line after which there are no points. This line indicates the self-imposed limitation of thermal conductivity of 1 W/m K. The line helps us understand the thermal diffusivity limitations given the thermal conductivity limitation for each reliability score. In real life, certain constraints on the parameters can be put, such as available materials' properties, to find other parameter distributions that would still satisfy the final system design constraints.

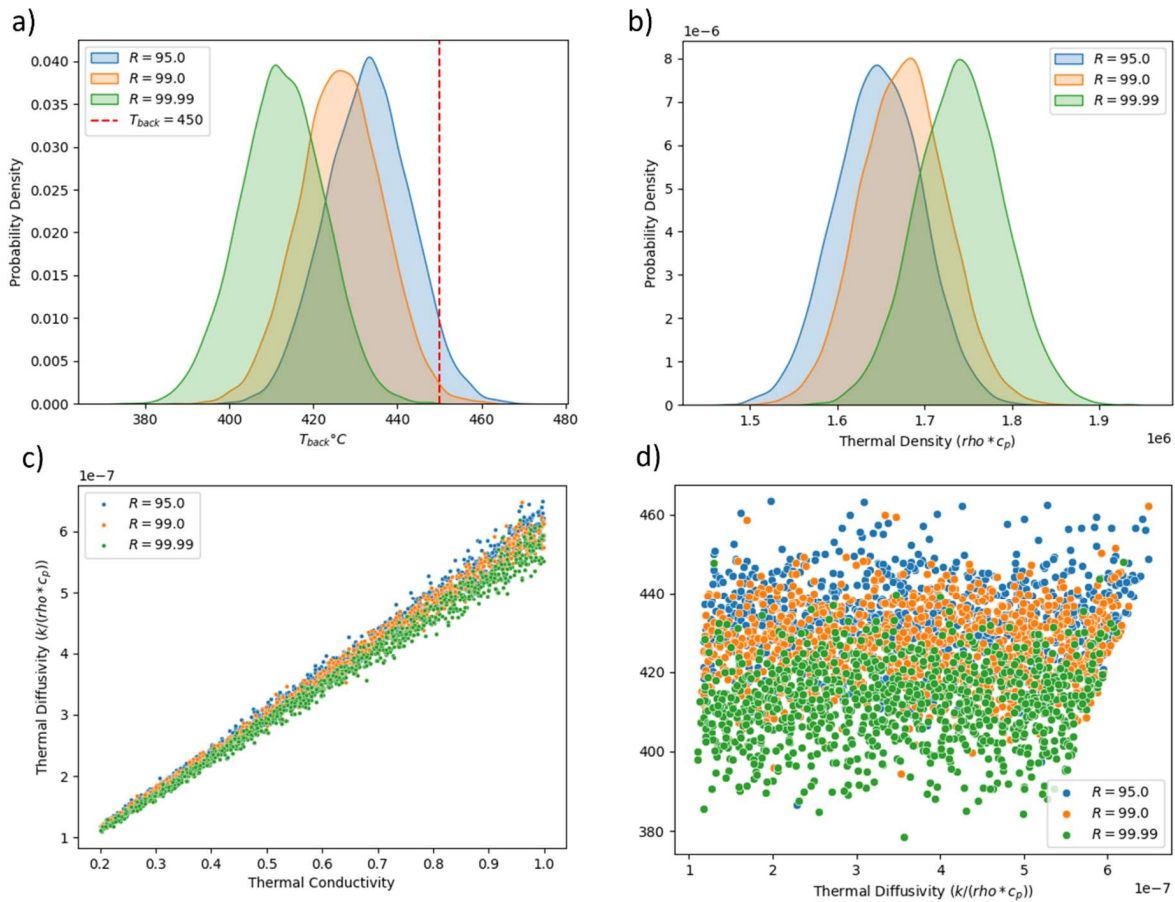

*Figure 8: a) Target Back Temperature Distributions with Different Reliability Scores b) Probability Density of Thermal Density ($\rho * c_p$) (c) Thermal Conductivity vs Thermal Diffusivity Distribution d) Thermal Diffusivity vs Back Temperature Distribution*

Now, the improvements in speed-ups reached from using PINN vs numerical model and SMC vs MCMC are compared. PINN inference time for a single simulation is about 9 times less than that of a numerical simulation. However, the difference becomes stark when the number of simulations are increased. The numerical solution times increase linearly for the number of simulations i.e. 2 simulations roughly take twice the time. But PINN, because of its parallelized nature, can generate results for hundreds of simulations at once. The time required to perform $N$ simulations is roughly the same as one if all the simulations can be done in parallel. Modern CPUs and especially GPUs have high memory capacity. Theoretically, millions of simulations can be solved at the same time as one. The time comparison for GPU is compared in the current study because for such small neural networks (0.089 MB was the network size), CPU can compete with the performance of a GPU. GPUs can be used for parallelization for more complex problems where larger neural networks are required.

SMC speed-up is quantified by running $N$ simulations in parallel. One simulation in parallel with one CPU means that it is essential performing MCMC, essentially no parallelization. The real speed-up from SMC comes when 100s of simulations are performed in parallel. The last point in figure 9b shows that compared to one simulation at a time as in MCMC, running all 10000 samples in parallel can speed-up the computation by 175 times. The whole process of generating target distribution, performing SMC sampling, adjusting the samples to get to the correct posterior distribution of parameters $\theta$ for 10,000 samples takes only 0.447 seconds. The same analysis performed using simple MCMC sampling and numerical solutions can take around 5-8 hours. This framework allows for practical implementation of material property constraints while maintaining system design requirements, offering significant computational efficiency through parallel processing capabilities.

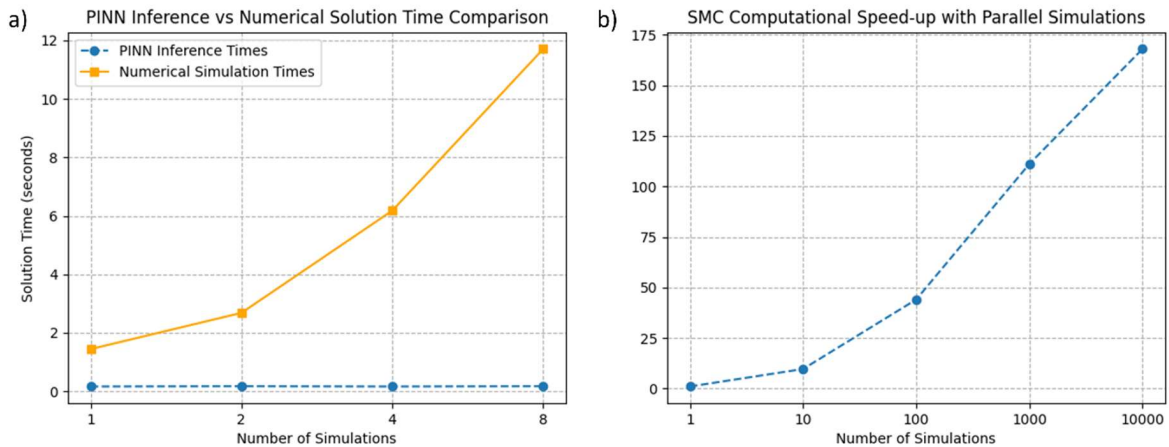

*Figure 9: a) PINN Inference Time vs Numerical Solution Time. Constant $\Delta t$ and $\Delta x$ are used for each numerical simulation for comparison b) SMC Computational Speed-up with Parallel Simulations*

# 6. Conclusion

This study investigated a new approach to designing thermal protection systems by combining physics-informed neural networks with sequential Monte Carlo sampling. The PINN model showed good accuracy while running many times faster for parallel simulations. Using SMC for parameter sampling provided substantial computational benefits, particularly when processing multiple simulations in parallel. When running 10,000 samples in parallel, the method achieved a 175x speed-up compared to traditional MCMC approach. The complete analysis process that previously took several hours using conventional methods was completed in under a second. The framework successfully handled different reliability requirements while considering material property constraints. These results suggest that combining PINN and SMC methods could be useful for improving the efficiency of thermal protection system design processes.